# Magnetization Enhancement in Magnetite Nanoparticles capped with Alginic Acid

B. Andrzejewski[a*], W. Bednarski[a], M. Kaźmierczak[a,b], K. Pogorzelec-Glaser[a], B. Hilczer[a],
S. Jurga[b], M. Matczak[a], B. Łęska[c], R. Pankiewicz[c], L. Kępiński[d]

[a]Institute of Molecular Physics, Polish Academy of Sciences
Smoluchowskiego 17, PL-60179 Poznań, Poland
* corresponding author: andrzejewski@ifmpan.poznan.pl

[b]NanoBioMedical Centre, Adam Mickiewicz University
Umultowska 85, PL-61614 Poznań, Poland

[c]Faculty of Chemistry, Adam Mickiewicz University
Umultowska 89b, PL-61614 Poznań, Poland

[d]Institute of Low Temperature and Structure Research, Polish Academy of Sciences
Okólna 2, PL-50422 Wrocław, Poland

*Abstract* — We report on the effect of organic acid capping on the behavior of magnetite nanoparticles. The nanoparticles of magnetite were obtained using microwave activated process, and the magnetic properties as well as the electron magnetic resonance behavior were studied for the $Fe_3O_4$ nanoparticles capped with alginic acid. The capped nanoparticles exhibit improved crystalline structure of the surface which leads to an enhanced magnetization. The saturation magnetization $M_s$ increases to ~75% of the bulk magnetization. The improved structure also facilitates quantization of spin-wave spectrum in the finite size nanoparticles and this in turn is responsible for unconventional behavior at low temperatures. The anomalies are apparent both in the electron magnetic resonance and in the magnetic properties. In magnetic resonance, they are manifested as an unusual increase in the resonant field $H_r(T)$ and also as a maximum of the spectroscopic splitting $g_{eff}$ parameter at low temperatures. The unconventional behavior of the nanoparticles is also responsible for a pronounced upturn of magnetization at low temperatures and a deviation from the Bloch law $M(T) \sim T^{1.5}$. At low temperatures, the magnetization loops $M(H)$ of the capped particles exhibits linear contribution related to an unconventional mechanisms of surface magnetism.

## I. INTRODUCTION

Magnetite $Fe_3O_4$ is one of the best and most preferred materials to combine with various polymers because of unique physical properties, like ferrimagnetic ordering, large magnetic moment, relatively high conductivity and high ratio of spin polarization. These features make magnetite a highly desired material for future applications in medicine [1], spintronics [2], electronic and optoelectronic devices [3-6], sensors [7] and for magnetic data storage [8].

Magnetite exhibits the inverse spinel $AB_2O_4$ crystallographic structure with two different kinds of voids in the *fcc* lattice of the $O^{2-}$ ions which are occupied by Fe atoms. Tetrahedral A voids contain $Fe^{2+}$ ions, whereas octahedral B voids are occupied by mixed valency $Fe^{2+}$ and $Fe^{3+}$ ions. Coupling between magnetic sublattices formed by the Fe moments located at sites A and B results in ferrimagnetic ordering. Mixed valence of the Fe ions and fast electron hopping between the B sites are responsible for a relatively high electric conductivity of $Fe_3O_4$ at high and moderate temperatures. However, below $T_v \approx 125K$ [9], the conductivity of magnetite decreases by two orders of magnitude. This transition has been for the first time observed and explained by Verwey in terms of electron localization in the octahedral B sublattice. Later, it has been found that only a small fraction of the electron population becomes localized below $T_v$ and the true nature of the Verwey transition is still under debate [9-11].

Nanostructured magnetite has distinct magnetic, electronic and optical properties when compared to those of the bulk material. Namely, an assembly of noninteracting nanoparticles exhibits superparamagnetism (SPM) at high temperatures and ferromagnetic (ferrimagnetic for magnetite) (FM) behavior below the blocking temperature $T_B$. Weak dipolar interactions between the nanoparticles lead to a supersipin glass (SSG) phase composed of large clusters of dipolar coupled nanoparticles. These clusters became blocked at low temperatures. Stronger interaction can even order the magnetic moments of individual clusters and induce superferromagnetic state (SFM) [12]. As regards the optical and electric properties, magnetite nanoparticles exhibit nonlinear absorption band of visible radiation and also photoinduced electric polarizability in weak optical fields [13]. The changes in polarizability are determined by intraband phototransition of nanoparticle charge carriers. Moreover, the optical properties of magnetite nanoparticles can be easily modified by an external magnetic field [14, 15]. It has been reported that the electric transport in 1D core-shell magnetite nanowires strongly depends on the applied magnetic field because of the tunneling magnetoresistance effect (TMR) [16, 17].

In high frequency microwave fields, the behavior of nanoparticles of iron oxides borders on classical thermodynamic and quantum dynamics and can be effectively studied by means of ferromagnetic (FMR) and electron magnetic resonance (EMR) [18]. FMR and the classical properties of an assembly of superparamagnetic nanoparticles are mainly observed in the high temperature limit. At low temperatures, on the other hand, EMR can be observed and described in terms of the resonance transitions summed over discrete energy spectrum of the nanoparticles. Quantization of the magnetic moment and surface anisotropy of individual particles are the another important factors that should be taken into account in this description.

Magnetite is also known as a stable, common and biocompatible mineral of very low toxicity (nanocrystals of magnetite are magnetoreceptors in some animal brains [19]). From this point of view, nanoparticles of magnetite offer large potential for biomedical applications, for example, in magnetic resonance imaging [20], magnetic targeting drug delivery [21] or as hyperthermia agents [22]. Unfortunately, significant reduction of magnetization occurs at the surface of $Fe_3O_4$ nanoparticles, which makes them useless for many of these applications. This obstacle can be overcame by capping the magnetic nanoparticles with various polymers [23-26] and organic acid, which in some cases allows restoration of the surface magnetism [27-29]. The best capping materials for medical applications are obviously organic ones like oleic and aliphatic acids [30], oleic acid with oleyamine [27], stearic acid [31], sulphamic acid [32], and many others [33]. One of the new capping materials is alginic acid – a cheap, common and nontoxic natural biopolymer [34-36] used already in food industry as a gelling agent or flavorless gum (sodium alginate E401).

The aim of this work is thus to yield information of the effect of alginic acid capping on the surface magnetization recovery and on the surface properties of $Fe_3O_4$ nanoparticles by means of EMR and magnetometric measurements.

## II. EXPERIMENTAL

*A. Sample Synthesis*

$Fe_3O_4$ magnetite nanoparticles were synthesized by means of microwave assisted hydrothermal reaction [37, 38] using: 1.623 g, 6.0 mM anhydrous iron(III) chloride $FeCl_3 \cdot 6H_2O$, 200 g, 87.8 mM sodium acetate $CH_3COONa$ and 0.420 g polyethylene glycol PEG 400 dissolved in 100 ml ethylene glycol. The solution was next stirred at room temperature until it changed the color to a homogeneous orange. After that, the mixture was transferred into a Teflon reactor (XP 1500, CEM Corp.) and loaded into a microwave oven (MARS 5, CEM Corp.). The reaction was carried out at 180°C for 30 min. After processing, the suspension of $Fe_3O_4$ nanoparticles was separated at 5000 rpm to obtain black precipitation. This precipitation was rinsed with anhydrous ethanol and vacuum dried at 60°C. To obtain composites of $Fe_3O_4$ nanoparticles with alginic acid (AA), water suspensions of magnetite nanoparticles and alginic acid were prepared and mixed together at the weight ratios of the magnetite to alginic acid ranging from 0.1 to 10%. The mixtures were processed for 15 min in an ultrasonic cleaner, poured into Petri dishes and air dried at room temperature. At the end of this procedure, fragile, dark in coloration, solid state flakes of $Fe_3O_4$-AA nanocomposite were obtained.

*B. Sample Characterization*

The crystallographic structures and phase compositions of the $Fe_3O_4$ nanopowders were studied by means of X-ray diffraction method (XRD) using an ISO DEBYEYE FLEX 3000 diffractometer fitted with a HZG4 goniometer in a Bragg-Brentano geometry and with a Co lamp ($\lambda$=0.17928 nm). The morphology and structure of $Fe_3O_4$ nanoparticles and $Fe_3O_4$-AA composites were studied by means of FEI NovaNanoSEM 230 scanning electron microscope (SEM) and also by Philips CM20 SuperTwin transmission electron microscope (TEM).

Differential Scanning Calorimetry (DSC) was used to study the thermal stability of the nanocomposites. DSC data were recorded with a Netzsch DSC-200 F3 calorimeter in the temperature range from 300 K to 500 K. Samples with a mass of approximately 3 mg were prepared in powder form. The measurements were carried out both on heating and cooling runs at a rate of 10 K/min.

Electron magnetic resonance studies of $Fe_3O_4$ nanoparticles and $Fe_3O_4$-AA composites were performed with a Bruker ElexSys E500 spectrometer operating in the S (~3.5 GHz) and the X-band (~9.5 GHz). S-band spectrometer was used only at room temperature to testify to a possible existence of narrow superparamagnetic components usually appearing above the blocking temperature $T_B$ [18]. Temperature measurements were performed by means of an X-band spectrometer fitted with Oxford Instruments and Bruker cryostats in the temperature ranges 4÷300 K and 300÷473 K, respectively. EMR spectra were recorded as the first derivative of the microwave power absorption vs. magnetic field in the range 0.005÷1.6 T. The magnetic field was modulated with a frequency of of 100 kHz and the amplitude of $10^{-4}$ T. EMR experiments were carried out with microwave power of about 2 mW. Magnetic measurements were performed using a Vibrating Sample Magnetometer (VSM) probe installed on the Quantum Design Physical Property Measurement System (PPMS). Temperature dependence of magnetization $M(T)$ were measured for the applied magnetic field 0.3 T.

Magnetization loops $M(H)$ were measured in the field range ±2 T for selected temperatures 10 K and 300 K.

## III. RESULTS AND DISCUSSION

TEM micrograph of the an assembly of as synthesized $Fe_3O_4$ nanoparticles is presented in Fig. 1. This assembly contains almost spherical, monodisperse nanoparticles. White spots appearing on the surface of some nanoparticles are concaves caused by local reduction of iron oxides. The reduction can take place during TEM examination in vacuum under high energy electron beam.

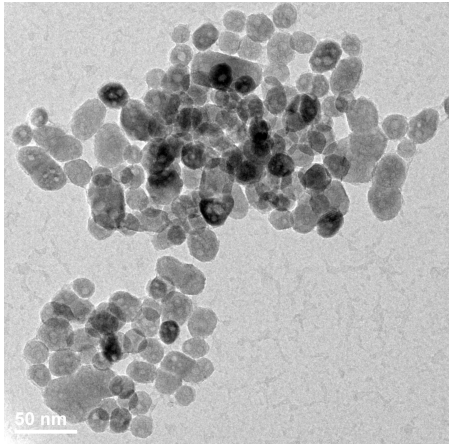

Figure 1.  The TEM micrograph of $Fe_3O_4$ nanoparticles.

The size of $Fe_3O_4$ nanoparticles varies from about 15 nm to 30 nm as shown in the histogram in Fig. 2. Most of the nanoparticles have diameter in the range 19÷22 nm. The size distribution of nanoparticles can be fitted with log-normal function commonly used to describe granular random systems [39]:

$$f(D) = \frac{1}{\sqrt{2\pi\sigma^2}} \frac{1}{D} \exp\left(-\frac{1}{2\sigma^2} \ln^2\left(\frac{D}{\langle D \rangle}\right)\right) \quad (1)$$

where $\langle D \rangle$ denotes the median size of the nanoparticles and $\sigma$ is the distribution width. The best fit of eq. (1) to the data presented in the histogram is obtained for the median size $\langle D \rangle$=20.6(1) nm and standard deviation $\sigma$=0.144(6).

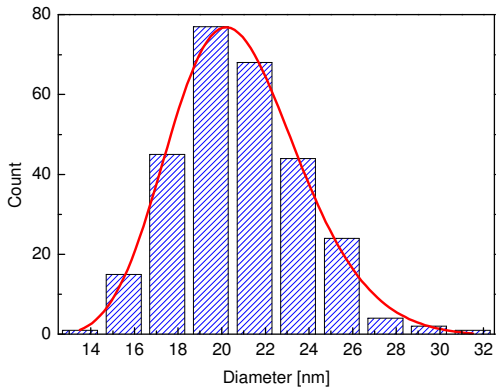

Figure 2.  The histogram of diameters of nanoparticles. The solid line is the best fit of the log-normal distribution eq. (1) to the data.

Crystallographic structure of the $Fe_3O_4$ nanoparticles was analyzed by means of HRTEM and a representative image is presented in Fig. 3. It shows that the nanoparticles are well developed magnetite crystallites, with almost no structural defects. The identified lattice fringes in Fig. 3 well correspond to the <111> planes of magnetite structure with a plane-to-plane separation of 0.486 nm and to <113> planes with a separation of 0.253 nm. A thin layer on the surface of the nanoparticles can be ascribed to amorphous $Fe_3O_4$ phase or a trace of resident surfactant (PEG 400) used in the synthesis.

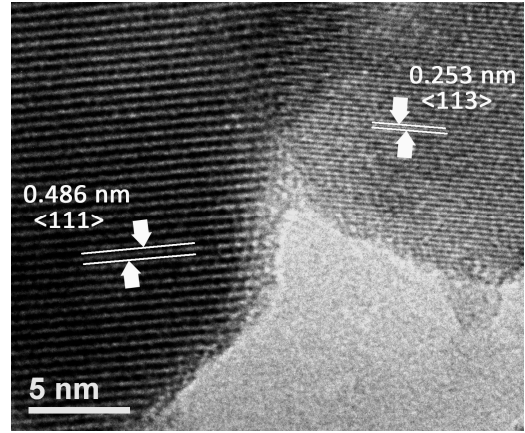

Figure 3.  The HRTEM micrograph of two selected $Fe_3O_4$ nanoparticles.

The selected area electron diffraction (SAED) pattern presented in Fig. 4 exhibits spotty diffraction rings and well resolved spots which confirm the single crystalline structure of the nanoparticles.

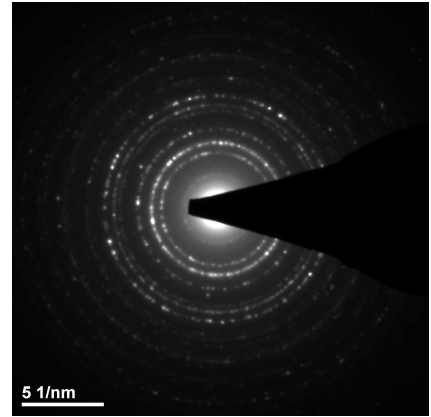

Figure 4.  The SAED pattern of $Fe_3O_4$ nanoparticle.

The crystallographic structure of $Fe_3O_4$ nanoparticles was verified also by means of X-ray diffraction performed at room temperature. The XRD patterns for as-prepared magnetite nanopowder, alginic acid and $Fe_3O_4$-AA composite are presented in Fig. 5. The solid lines in Fig. 5a correspond to the best fits obtained by means of Rietveld method (calculated using FULLPROF software) to the experimental data. The lines below the XRD data illustrate the difference between the data and the fit. The vertical markers represent the calculated positions of Bragg peaks. Analysis of these XRD data indicates the presence of Fe$_3$O$_4$ phase with cubic structure and *Fd3m* space group. The parameters of the cubic unit cell are as follows: *a*=8.374(2) Å and *α*=*β*=*γ*=90.

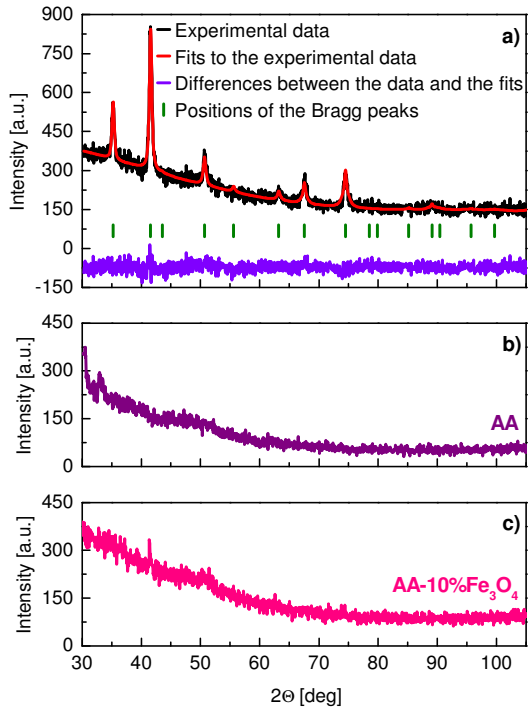

Figure 5. The XRD-pattern of Fe$_3$O$_4$ nanoparticles (a), alginic acid (b) and Fe$_3$O$_4$-AA composite containing 10 wt. % of magnetite (c). Solid lines are the best fits to the experimental data. The lines below the fits represent the differences between the data and the fits. The vertical markers represent positions of the Bragg peaks.

The mean size of Fe$_3$O$_4$ nanocrystallites was estimated using the Scherrer's formula [40]: $D_S = K\lambda/\beta\cos\Theta$, where $D_S$ denotes the crystallite size, $\beta$ is the half-width of the diffraction peak and $\Theta$ represents the position of the Bragg peak. The constant K in the Scherrer's equation [40] depends on the morphology of the crystallites and K≈1 is usually assumed. The mean size of the crystallites determined from XRD measurements is comparable with that obtained from TEM micrographs and equals 18.3±0.7 nm. The reason for this small discrepancy can be an amorphous layer or a layer of PEG 400 surfactant on the surface of Fe$_3$O$_4$ nanocrystals, which can lead to overestimation of the nanoparticle sizes in TEM micrographs. We do not expect that this disagreement can be due the internal strain causing additional broadening of the diffraction peaks, which would lead to underestimation of the nanocrystal size when using the Scherrer's equation. The hydrothermally synthesized powders are usually free of strain. Indeed, the analysis of the internal strain $\varepsilon$ by means of Williamson formula $\beta = C\varepsilon\tan\Theta + K\lambda/D_S\cos\Theta$ [41], (it is assumed that the constant C≈4) has shown a negligible amplitude of the strain $\varepsilon \approx 2\cdot 10^{-5}$. The XRD spectrum for alginic acid in Fig. 5b does not exhibit any Bragg peaks in the reported 2$\theta$ range. The Fe$_3$O$_4$ nanoparticles characterized above in details and alginic acid were used to synthesize Fe$_3$O$_4$-AA nanocomposites. The XRD data in Fig. 5c for Fe$_3$O$_4$-AA exhibit only two very weak maxima at 41$^0$ and at about 50$^0$ originating from Fe$_3$O$_4$, even at the highest load of magnetite in the manufactured set of composites equal to 10 wt.%. The intensity of these peaks is too low to perform any detailed analysis.

The SEM micrograph of the nanocomposite, which contains 0.5 wt. % of magnetite, is shown in Fig. 6. The structures of the other nanocomposites with different load of magnetite are very similar. Because of small magnification, such details of the structure as individual nanoparticles are not visible. Examinations of the structure at higher magnifications were difficult because of very high resistivity of the sample.

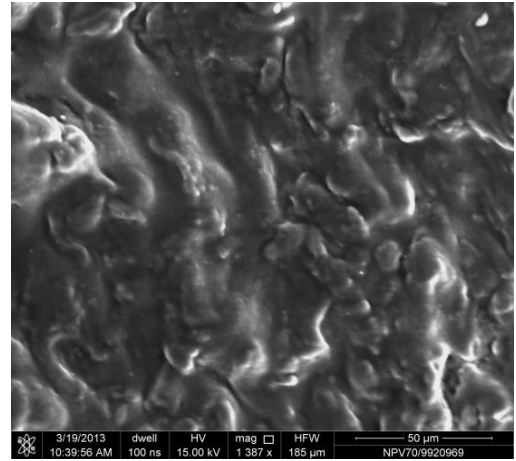

Figure 6. SEM micrograph of Fe$_3$O$_4$-AA nanocomposite with the content of magnetite equal to 0.5 wt. %.

The results of measurement of the Fe$_3$O$_4$-AA nanocomposite containing 0.1 wt. % of magnetite, by means of DSC method are presented in Fig. 7. The DSC curve exhibits endothermic maxima at about 360.2 K, 441.2 K and 448.5 K.

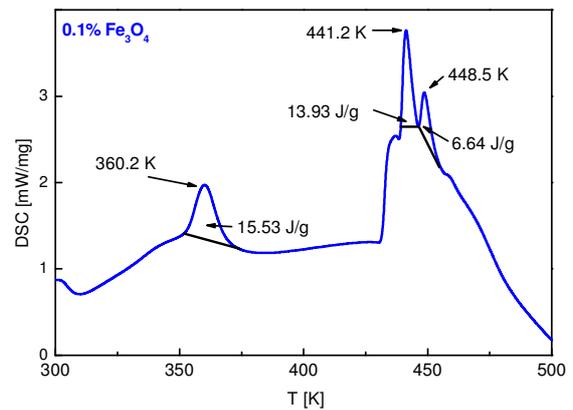

Figure 7. DSC curve for the Fe$_3$O$_4$-AA nanocomposite containing 0.1 wt. % of magnetite; $dT/dt$=10 K/min.

The low temperature maximum can be related to the removal of absorbed water, whereas the two maxima at higher temperatures are assigned to decomposition and melting of the nanocomposite. Decomposition of the nanocomposite above 441.2 K is also seen in EMR spectra as disappearance of the additional free radical signal.

As reported by Durmus et al. and Unal et al. [34-36], alginic acid in the composite is not simply adsorbed at the surface of nanoparticles but rather bonded to it via bridging oxygens of the carboxylate groups. This chemical bonding can modify the surface, the magnetocrystalline anisotropy of the nanoparticles and also the oxidation state of iron in the magnetite. To study these effects we used EMR and magnetometric measurements.

The EMR spectra of the samples recorded in the S and X microwave bands and presented in Fig. 8, consist of show single, nonsymmetrical, broad lines with shapes that suggest cubic (negative) or uniaxial (negative) magnetic anisotropy [42]. The lack of a narrow line in the spectra at the $g$-factor $g≈2$ (up to the temperature of alginic acid decomposition and free radical generation) does not indicate nanoparticle superparamagnetic behaviour above the blocking temperature $T_B$ [43, 44] for both microwave frequencies and applied DC magnetic fields. This is consistent with our expectations, because the narrow line is usually recorded for isolated, smaller nanoparticles with a size less than 10 nm [43, 44]. Formation of the chains with aligned dipole moments [45] of nonisolated $Fe_3O_4$ nanoparticles causes a narrowing of the EMR signal in comparison with the signal assigned to dispersed nanoparticles functionalized with alginic acid as shown in Fig.8 .

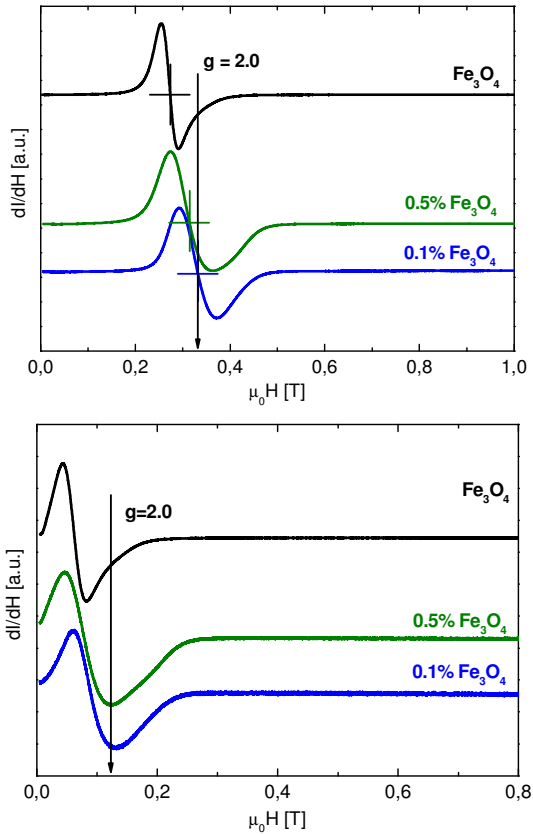

Figure 8. a) X–band (9.5 GHz) and b) S-band (3.5 GHz) EMR spectra of $Fe_3O_4$ nanoparticles and nanoparticles in AA matrix recorded at 300 K.

We cannot also exclude an exchange interaction between magnetite nanoparticle interfaces which should lead to a strong narrowing of EMR line width. On the other hand, increase in the linewidth broadening with the magnetite concentration in alginic acid (compare the EMR spectra recorded for 0.1 wt.% and 0.5 wt. % $Fe_3O_4$) can be due to enhancement of the dipole-dipole interaction. This broadening of the line width can be also caused by an increased disorder at the surface of AA functionalized magnetite nanoparticles, because the chemical bonds between acid carboxylic groups an Fe ions are formed at certain surface sites, only [46]. In general, a significant increase in the EMR linewidth $\Delta H_{pp}$ (Fig. 9) with decreasing temperature is a common feature of nanoparticles exhibiting magnetic moment [43, 44, 47].

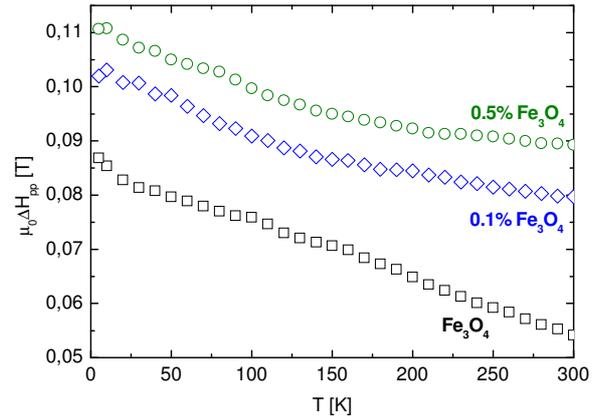

Figure 9. Temperature dependences of X-band EMR linewidth $\Delta H_{pp}$ of interacting magnetite bare nanoparticles and capped nanoparticles.

For the nanoparticles studied one can observe a shift of the EMR line positions $H_r$ (defined as the field for which $dI/dH$=0 and marked by crosses in Fig.8) to the lower magnetic field upon cooling as shown in Fig. 10. This type of temperature variation of the line width $\Delta H_{pp}$ and line positions $H_r$ in the EMR spectra was attributed to a gradual suppression of the averaging effect of thermal fluctuations with decreasing temperature [47-49].

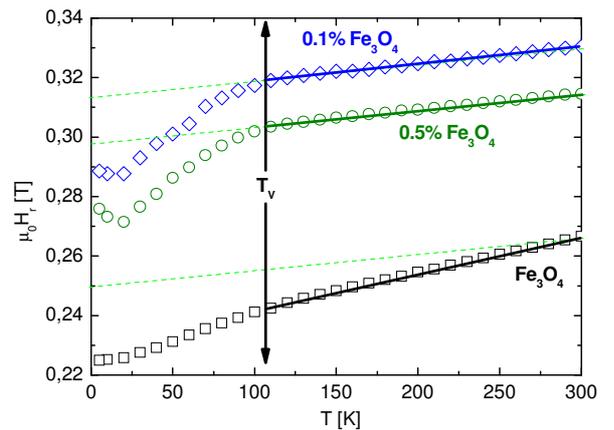

Figure 10. Temperature dependences of X-band resonance line positions. Solid lines present the fits with equation (1), dashed lines indicate linear functions with the same slopes *i.e.* $k_{eff}/M_s$=const.

According to the simple model described in [50], temperature variations in $H_r$ of a spherical magnetic nanoparticle can be approximated by a linear function:

$$H_r = \left(\frac{\omega_R}{\gamma} - H_D - \frac{2K_B + K_S^0}{M_S}\right) + \frac{2k_{eff}}{M_S}T \quad (2)$$

where $\omega_R$ and $\gamma$ are the Larmor frequency and the gyromagnetic ratio, respectively, $H_D$ denotes the demagnetization field, $K_B$ and $K_S^0$ are the magnetocrystalline anisotropy density and the surface anisotropy density at temperature $T=0$ K, $M_S$ is the saturation magnetization and $k_{eff}$ is a coefficient related to temperature changes in the effective magnetic anisotropy.

Fig. 10 shows, that eq. (2) describes correctly the experimental data in the temperature range 300-115 K. The change in the temperature derivative of the magnetic resonance field near 115 K reveals the Verwey transition. It is also evident that in the temperature range 300-115 K, the slope $k_{eff}/M_s$ of the $H_r(T)$ line, has the same value for various concentrations of magnetic nanoparticles in $Fe_3O_4$-AA samples in contrast to the interacting bare magnetite nanoparticles in $Fe_3O_4$ sample. Lower value of the $k_{eff}/M_s$ ratio for $Fe_3O_4$-AA may suggest significant changes in the $k_{eff}$ coefficient and/or enhancement of the saturation magnetization $M_s$, due to the modification of the nanoparticle surface by alginic acid. Increase in the value of the first term of eq. (1) indicates weakening of dipolar interactions $H_D$ as the concentration of the nanoparticles decreases. Note also that the Verwey transition is much more pronounced in the capped $Fe_3O_4$ nanoparticles than in the bare ones. At the lowest temperature we observe an unusual increase in the resonant field $H_r(T)$ presented in Fig. 10. This feature seems to be more apparent for the nanocomposites than for $Fe_3O_4$ nanoparticle assembly, which confirms our earlier assumption that the chemical bonding between AA and magnetite modifies the nanoparticles properties. In low temperatures, $g_{eff}$ parameter exhibits also unusual maximum at about 15 K (Fig. 11). Moreover at a constant temperature, $g_{eff}$ values are different for cooling and heating cycles. This suggests nonergodic state of magnetite nanoparticles below 15 K.

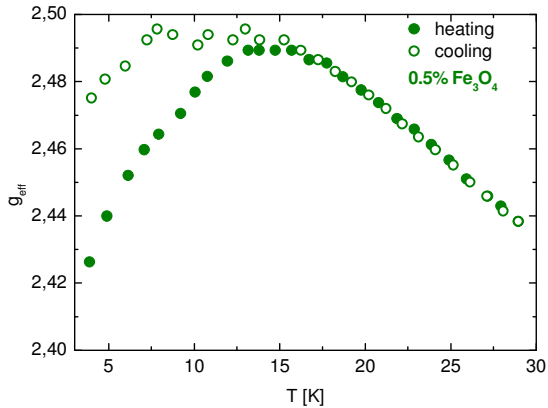

Figure 11. Temperature dependences of $g_{eff}$ parameter. The cooling and heating cycles are represented by open and solid circles, respectively.

Low temperature anomaly near 25÷30 K was observed in monocrystalline magnetite and thin films for magnetic aftereffect spectroscopy [9] and ac susceptibility measurements [51], respectively. This anomaly was assigned to incoherent electron tunneling [9]. In the present studies, the low temperature anomaly takes place at lower temperature and seems to have another origin. As shown in [52] the surface effects are even more pronounced for magnetic nanoparticles than in thin films and because of this, the surface of magnetite nanoparticles can be strongly influenced by bonding to the alginic acid. The chemical bonding of organic acid chains to the nanoparticles surface takes place via oxygen from the carboxylic groups. This allows a restoration of the local environment of surface iron atoms close to bulk $Fe_3O_4$ crystal coordination. Therefore the nanoparticles of magnetite embedded in alginic acid exhibit much improved crystalline structure and the quantum effects, common for small size objects, can appear at low temperatures. We assume that the low temperature anomalies in $H_r$ and $g_{eff}$ can be considered as related to a quantization of spin-wave spectrum in the finite size nanoparticles. In bulk ferro- or ferrimagnets the spin wave excitation spectrum can be treated as continuous with the partition $Z=\int g(k)n(k)dk$ where $g(k)$ is the density of spin wave states and $n(k)$ is the occupation number. However, in small nanoparticles, long spin–waves cannot propagate, the spectrum becomes discrete at low temperatures and the partition of the states is now the sum $Z=\sum 1/[\exp(E_n/k_BT-1)]$, where $k_B$ is the Boltzmann constant. The changes in the spin-wave spectrum alter the populations of magnetic sublevels $\rho'_{m,\theta}$ with energy $E_{m,\theta}$ because they are dependent on $Z$: $\rho'_{m,\theta}=\exp(-E_{m,\theta}/k_BT)/Z$. This influences the shape, intensity $I$ and parameters of the EMR signal originating from the assembly of nanoparticles with anisotropy axes directed at various $\theta$ angles with respect to the external magnetic field $H$ [18]:

$$I(H-H_r) = \int_0^\pi \sin\theta d\theta \int_{-S}^S W_{m,\theta}\frac{\partial \rho_{m,\theta}}{\partial m}dm \quad (3)$$

In eq. (3) the probability of the allowed transitions $W_{m,\theta}=Ag[S(S+1)-m(m+1)]$ depends on the proportionality coefficient A, the total spin $S$ of the nanoparticle, magnetic quantum number $m$ and the form-factor function $g(H-H_{m,\theta})$ of the resonance line.

Quantization of the spin-wave modes can be also observed in magnetic properties of $Fe_3O_4$-AA composites and it is manifested in the upturn of the magnetization at low temperatures, shown in Fig. 12. The increase in magnetization to the low temperature value $M_s(0)$, is due to the excitation of the spin waves with discrete spectrum [53] and can be fitted with the equation:

$$M_S(T) = M_S(0) - C\left[\exp\left(-\frac{E_1}{k_BT}\right) + \exp\left(-\frac{E_2}{k_BT}\right)\right] \quad (4)$$

where $C=M_s(0)/Nn$ is the coefficient dependent on the number of the modes $N$ and their occupancy $n$.

The best fit of eq. (4) to the low temperature upturn of magnetization can be obtained for the following parameters: $M_s(0)=119.6\pm0.2$ Am$^2$/kg, $C=37.2\pm0.2$ Am$^2$/kg; $E_1/k_B=2.5\pm0.5$ K, $E_2/k_B=200\pm20$ K for the content of magnetite 0.1 wt. % and $M_s(0)=81\pm0.2$ Am$^2$/kg, $C=14.5\pm0.1$ Am$^2$/kg; $E_1/k_B=2.5\pm0.5$ K, $E_2/k_B=140\pm20$ K for 0.5 wt. % of magnetite. At higher temperatures, the magnetic energy levels are broadened and form a continuous excitation spectrum. In this case, the dependence of magnetization of Fe$_3$O$_4$-AA composites on temperature can be described, like in bulk materials, by the Bloch law: $M_s(T)= M_s(0)(1-BT^{3/2})$ where B is a coefficient related to the atomic volume and the spin-wave stiffness. The best fits are obtained for the parameters: $M_s(0)=85.5\pm0.5$ Am$^2$/kg, $B=4\cdot10^{-5}\pm1\cdot10^{-6}$ K$^{-3/2}$ for 0.1 wt. % of Fe$_3$O$_4$ in nanocomposite, $M_s(0)=67.4\pm0.5$ Am$^2$/kg, $B=3.1\cdot10^{-5}\pm1\cdot10^{-6}$ K$^{-3/2}$ for the nanocomposite containing 0.5 wt.% of magnetite. The Bloch law, however, is invalid for the uncapped Fe$_3$O$_4$ nanoparticles which exhibit different temperature dependence of magnetization $M(T)\sim T^{1.9}$. The deviation from the Bloch law can be related to the degraded surface with reduced magnetization and/or to the dipolar or exchange interactions in the dense assembly of Fe$_3$O$_4$ nanoparticles. Also, there is no anomaly at low temperatures, which indicates that spin-waves excitations are dumped in bare nanoparticles.

The Verwey transition does not appear in magnetic measurements as it had been already noticed in many earlier reports [34, 49, 54, 55]. The lack of the Verwey transition happens if the magnetite nanoparticles are in the superparamagnetic state *i.e.* if the blocking temperature $T_B$ of thermally activated reorientations of magnetization is higher than the Verwey temperature $T_v$. The blocking temperature is related to the time of the measurement $\tau_m$

$$T_B = \frac{K_B V}{k_B} \Big/ \ln\frac{\tau_m}{\tau_0} \qquad (5)$$

where $V$ is the volume of nanoparticle and $\tau_0\approx10^{-10}$.

For the S (or X) microwave band $\tau_m\approx2.9\cdot10^{-10}$ s ($\approx1.1\cdot10^{-10}$ s), which gives $T_B\approx K_B V/k_B$ ($\approx10K_BV/k_B$). In the case of magnetic study the time of measurement is much longer and equal $\tau_m\approx1$ s. Consequently, the blocking temperature $T_B$ for magnetic measurements is about 20 times lower than that for EMR. Thus nanoparticles can still exhibit superparamagnetic behavior, whereas at the same applied magnetic field during microwave measurements they are in FM and nonergodic state. This nonergodic, irreversible behavior is apparent in Fig. 11. On the other hand, no difference between zero field cooling (ZFC) and field cooling (FC) magnetization curves indicates superparamagnetic, reversible properties and a very low blocking temperature.

If the magnetization is determined with respect to the mass of the magnetite, as in Fig. 12, it turns out that the composites exhibit enhanced magnetic properties compared to that of the bare magnetite nanoparticles. At room temperature, the saturation magnetization is $67.6\pm0.1$ Am$^2$/kg for the composite with magnetite content 0.1 wt. % the bulk magnetization of magnetite is ~92 Am$^2$/kg, $56.5\pm0.1$ Am$^2$/kg for the nanocomposite containing 0.5 wt. % of Fe$_3$O$_4$ and only $47.5\pm0.1$ Am$^2$/kg for bare magnetite nanoparticles.

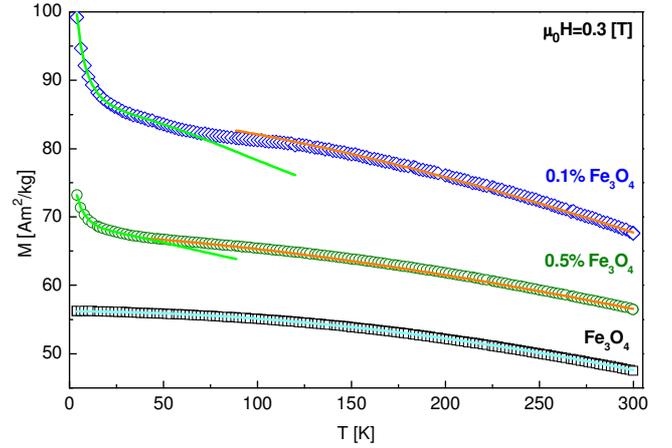

Figure 12. Temperature dependences of magnetization of Fe$_3$O$_4$ nanoparticles and of Fe$_3$O$_4$-AA composites. The magnetization was calculated with respect to the mass of magnetite. The applied magnetic field was 0.3 T.

The reduced magnetization at the bare surface results from depleted number of nearest neighbors oxygen atoms surrounding the octahedral iron, which is five instead of six as in the bulk magnetite. Also the surface in-plane oxygens ions are closer to the Fe ions, increasing the hybridization of $d_{x^2-y^2}$ orbitals. The orbitals become partially empty as they are shifted away from the Fermi level, which leads to a decay of magnetic moment at the surface of the magnetite. On the other hand, the alginic acid capped nanoparticles exhibit improved surface due to the chemical bonding between alginic acid and the surface of magnetite. More exactly, these chemical bonds are formed between the oxygen atoms from carboxylic groups of the acid and the two of four Fe ions at surface unit cell of the magnetite, whereas the rest of Fe ions remain unbonded [46]. This allows supplementing to six the number of oxygen atoms of the octahedrally coordinated surface ions of iron. Also, the occupancies of $d$ orbitals are back to bulk Fe$_3$O$_4$ crystal value. In effect, the magnetic moment of the capped nanoparticles is 1 Bohr magneton per surface unit cell larger than that of the bare surface of magnetite [46]. On the capped surface of magnetite there is a chessboard of bonded and unbonded Fe ions and thus one can expect unconventional magnetic orderings. Indeed, unconventional behavior of magnetization occurs in magnetization loops recorded for Fe$_3$O$_4$-AA composites at low temperatures presented in Fig. 13. Both the bare Fe$_3$O$_4$ nanoparticles and Fe$_3$O$_4$-AA composites exhibit FM hysteresis loops that saturate above about 0.3 T. Once again, the magnetization of the composites is enhanced as compared to that of the bare Fe$_3$O$_4$ nanoparticles because of bonding to alginic acid. Negative contribution to magnetization, observed at high field for the composite with Fe$_3$O$_4$ load 0.1 wt. % originates from the alginic acid matrix. However, at low temperatures the magnetization loops for the composites are superpositions of FM and positive, linear contributions at high fields from unconventional magnetic ordering.

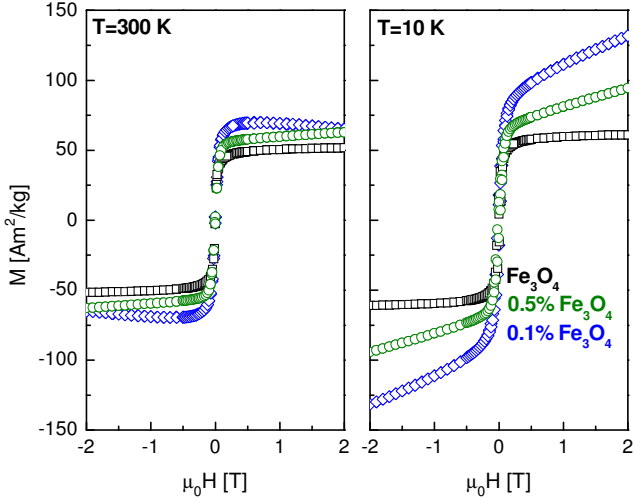

Figure 13. Magnetization loops of $Fe_3O_4$ bare nanoparticles and $Fe_3O_4$-AA composites recorded at room temperature and at 10K. The magnetization was normalized with respect to the mass of magnetite.

This unconventional contribution cannot be simply related to the paramagnetic moment from the unbonded and disordered $Fe_3O_4$ regions, because it is completely absent in the bare nanoparticles with highly degraded surface. Moreover, the linear contribution occurs at low temperature only, where the spectrum of spin-wave excitation is discrete. It may indicate a coupling between spin-wave modes and surface spins located in the alternating regions which can be degraded or bonded to alginic acid.

## IV. CONCLUSIONS

In summary, we have performed a comparative study of bare magnetite nanoparticles obtained by means of microwave activated process and of the same particles capped with alginic acid. The EMR spectra of bare and capped nanoparticles both indicate FM behavior with negative cubic or uniaxial magnetic anisotropy. The lack of a narrow line in the EMR spectra at $g \approx 2$ indicates the absence of nanoparticle superparamagnetic behavior. The linear variation of the EMR signal position $H_r(T)$ on temperature can be well described using a simple model of the spherical magnetic nanoparticles exhibiting surface and magnetocrystalline anisotropy. The slopes $k_{eff}/M_s$ of the $H_r(T)$ lines in the case of the $Fe_3O_4$-AA composites are less steep than for the assembly of bare $Fe_3O_4$ nanoparticles and indicate a substantial increase in saturation magnetization $M_S$ for capped nanoparticles. The change in the slope $H_r(T)$ lines at 115 K reveals the Verwey transition. In the capped nanoparticles and at the lowest temperature of about 15K, an unusual increase in the resonant field $H_r(T)$ and a maximum in $g_{eff}$ parameter of the EMR signal take place.

Also magnetization $M(T)$ of capped nanoparticles exhibits an unusual behavior as at low temperatures it strongly deviates from the Bloch law $M(T) \sim T^{1.5}$. These unusual low temperature anomalies detected in EMR and in magnetic study we explain in terms of spin-wave spectrum quantization in the finite size nanoparticles. The quantization can take place in capped nanoparticles only, because they exhibit improved crystalline structure of the surface. More exactly, the capping with the alginic acid, allows supplementing the number of oxygens and restoring the local coordination of the surface ions of iron to those of the bulk $Fe_3O_4$ crystal environment. The capping also recovers the surface magnetization of the nanoparticles. This enhanced magnetization of nanoparticles is well manifested in the magnetization hysteresis loops $M(H)$. At low temperatures, where the spectrum of spin-wave excitation becomes discrete, the magnetization loops for the capped particles are superpositions of FM and linear contributions. The latter one may indicate unconventional mechanisms of surface magnetism in $Fe_3O_4$-AA composites.

## V. ACKNOWLEDGEEMENTS

This project has been supported by National Science Centre (project No. N N507 229040). M.K. was supported through the European Union - European Social Fund and Human Capital - National Cohesion Strategy.